# Site-selective mapping of metastable states using electron-beam induced luminescence microscopy

R. Kumar[1,*], L.I.D.J. Martin[2], D. Poelman[2], D. Vandenberghe[3], J.De Grave[3], M. Kook[1], M. Jain[1]

[1]Department of Physics, Technical University of Denmark, DTU Risø Campus, Denmark-4000

[2]Department of Solid-state Sciences, University of Ghent, Belgium-9000

[3]Department of Geology, University of Ghent, Belgium-9000

*Corresponding author: Raju Kumar (mail id: rajcelos.isp65@gmail.com)

Metastable states created by electron or hole capture in crystal defects are widely used in dosimetry and photonic applications. Feldspar, the most abundant mineral in the Earth's crust (>50%), generates metastable states with lifetimes of millions of years upon exposure to ionizing radiation. Although feldspar is widely used in dosimetry and geochronometry, the creation of metastable states and charge transfer across them is poorly understood. Understanding such phenomena requires next-generation methods based on high-resolution, site-selective probing of the metastable states. Recent studies using site-selective techniques such as photoluminescence (PL), and radioluminescence (RL) at 7 K have revealed that feldspar exhibits two near-infrared (NIR) emission bands peaking at 880 nm and 955 nm, which are believed to arise from the principal electron-trapping states.

Here, we map for the first time the electron-trapping states in potassium-rich feldspar using spectrally-resolved cathodoluminescence microscopy at a spatial resolution of ~6 to 22 μm. Each pixel probed by a scanning electron microscope provides us a cathodoluminescence spectrum (SEM-CL) in the range 600-1000 nm, and elemental data from energy-dispersive x-ray (EDX) spectroscopy. We conclude that the two NIR emissions are spatially variable and, therefore, originate from different sites. This conclusion contradicts the existing model that the two emissions arise from two different excited states of a principal trap. Moreover, we are able to link the individual NIR emission peaks with the geochemical variations (K, Na and Fe concentration), and propose a cluster model that explains the quenching of the NIR emission by $Fe^{4+}$. Our study contributes to an improved understanding of charge storage in feldspathic minerals, with implications for developing sub-single grain (micrometer scale) measurement techniques in radiation dosimetry.

# 1. Introduction

Feldspar, a natural crystalline aluminosilicate, is the most abundant mineral in the Earth's crust (> 50%). It crystallizes in monoclinic or triclinic systems depending on the temperature of formation. The crystal structure consists of $AlO_4$ and $SiO_4$ tetrahedrons connected through shared oxygen atoms forming a three dimensional (3D) framework. In the framework, Si and Al sites are named as the T sites, while the interstices (normally occupied by cations K, Na, or Ca) are called the M sites. Based on whether M = K, Na or Ca, a feldspar crystal occurs as $KAlSi_3O_8$ (sanidine, orthoclase, and microcline), $NaAlSi_3O_8$ (albite), or $CaAl_2Si_2O_8$ (anorthite); these are end members of the K-Na (alkali feldspar) and Na-Ca (plagioclase) solid solution series. Furthermore, there exist $T_1$ and $T_2$ sites because of two distinct Si or Al sites lying in different environments[1,2].

Luminescence from feldspar arises mainly due to defects and impurities that occupy energy levels within the bandgap (bandgap, $E_g \sim 7.7$ eV)[3]. For example, well-known emissions occur from the $Fe^{3+}$ and $Mn^{2+}$ ions occupying the $T_1$ and M sites, respectively[1,2]. Some defects form metastable states by capturing free electrons or holes created due to ionizing radiation. Annihilation of these states can be induced by thermal or optical stimulation; this leads to electron-hole recombination and subsequent luminescence emission in the ultra-violet (UV) to near-infrared (NIR) range. The fact that the resulting luminescence intensity is proportional to the prior concentration of the metastable states is exploited in solid-state luminescence dosimetry. Here the change in the concentration of the metastable states due to absorbed dose (unit Gy= J $Kg^{-1}$) from ionizing radiation is read out via luminescence[4]. The most widespread dosimetric application of feldspars is in luminescence dating, which uses the time-dependent accumulation of dose from environmental ionizing radiation to establish sediment deposition chronologies[5–10].

Radiation dosimetry using feldspar is based on the main electron-trapping defect called the principal trap. Electrons in the principal trap can be excited using near-infrared light (850 ± 30 nm), and the luminescence arising from electron-hole recombination is measured in the violet-blue emission band (340-450 nm). This technique called infrared stimulated luminescence[11] (IRSL) is a variant of the optically stimulated luminescence (OSL) method.

Despite rapid instrumental and methodological advances over the last two decades[1,7,12–14], the basic characterization and the exact identification of the metastable states giving rise to OSL or IRSL have not been possible. This lack of knowledge can be partly attributed to the fact that OSL (or IRSL) is not an ideal method to study the physical characteristics of the traps (e.g., excited state energies, trap depth, number of traps and distribution) since the luminescence signal is a convolution of electron excitation, transport, and recombination.

In contrast to OSL, infrared radioluminescence[15,16] (IRRL) and infrared photoluminescence[17,18] (IRPL) arise directly from the electron traps without involving any electron-hole recombination. IRRL provides real-time measurement of luminescence emitted during the capture of electrons in the principal trap while feldspar is exposed to ionizing radiation. IRPL measures electrons trapped in the principal trap after exposure to ionizing radiation using the photoluminescence (PL) technique; electrons are excited with NIR and the resulting Stokes-shifted emission from the relaxation of the excited state is measured[17].

Previous studies show the existence of two overlapping emissions in both IRRL and IRPL emission spectra. In the case of IRRL, emissions inferred to be centered at ~865 and ~910 nm (based on spectral deconvolution) have been attributed to two excited states of a $Pb^+$ electron trapping center by Erfurt [20]. Based on low-temperature spectroscopy, Kumar et al.[18] recently reported two distinct emission bands centered at ~880 nm and ~955 nm, both in IRRL and IRPL at 7 K. The authors confirmed experimentally that IRRL and IRPL are two complementary signals from the same trap. Based on the temperature and sample dependent variations in the two IRPL bands (~880 nm and ~955 nm), Kumar et al.[18] suggested that these emissions may arise from the same defect residing in two different sites (i.e. two principal traps) , and not the two excited energy states of the same site as inferred by Erfurt[20]. In order to better understand the characteristics of the electron trapping centers in feldspars, it is important to establish whether the two NIR emissions (~880 nm and ~950 nm) arise from the same defect or two different defects, how these defects are distributed in the crystal, and what their link is to the feldspar's mineralogical composition. This knowledge has important implications for improving the accuracy and precision in radiation dosimetry and luminescence dating, especially if the two emissions have different stabilities and dosimetric characteristics.

Since both IRPL and IRRL techniques examine the trapped electrons, selectively, they have the potential for site-specific probing of the principal trap. One way to examine the link between the emission band and the trapping state is to map the emissions spatially and understand their linkages to the chemical composition. Recently, Thomsen et al.[21] investigated the spatial distribution of IRPL in feldspar and related this to the corresponding map of the chemical composition obtained using micro x-ray fluorescence (μ-XRF). These authors concluded that IRPL mainly arises from the K-rich regions. The available spatial resolution in this study (because of cross-talk) restricted interpretations at the sub-grain level, and furthermore, only one IRPL emission (955 nm emission) was examined. Similarly, Sellwood et al.[22] have shown that IRPL (955 nm emission) is correlated with K concentration in feldspar in granitic rocks.

This work aims at improving our understanding of the behavior of metastable states in feldspar, the most widespread naturally occurring dosimeter. Such an understanding is essential for developing next-generation luminescence dating techniques at sub-single-grain spatial resolution. We use cathodoluminescence (CL; luminescence emitted during interaction of an electron beam with the sample) microscopy combined with visible-NIR spectroscopy[23,24] to map the two NIR emissions (880 nm and 955 nm), simultaneously with energy-dispersive x-ray spectrometry to determine elemental composition. This is the first investigation studying the spectrally-resolved spatial distribution of the principal trap emissions (880nm and 955nm) and its link to feldspar geochemistry.

## 2. Previous studies on cathodoluminescence in feldspar

Cathodoluminescence (CL) is emitted when a sample is exposed to an electron beam. The mechanism is similar to radioluminescence which is induced by ionizing radiation (x-rays, ϒ-rays, and α and β particles). The main difference is in the penetration depth of the different particles; CL typically samples the near-surface volume of the material, of the order of 1 μm, depending on the acceleration voltage of the electron beam and the sample composition. There are several ways by which CL in semiconductors or insulators is produced[25–27]: 1) band-to-band transitions, 2) band- to-defect transitions i.e. trapping of charge by a defect, 3) trap-assisted, band-to-band transitions, and 4) intra-defect transitions. Cathodoluminescence of feldspars is an important tool for interpreting genetic conditions of rock formation and alteration[2,25]. Forensic science makes use of CL to study the provenance of feldspar[28]. CL has also been used to study their luminescence behavior, and identify activators or defects[2,25]. These investigations have mainly focused on greenish-yellow (~560 nm) and red (~700 nm) emission bands that are suggested to arise from $Mn^{2+}$ and $Fe^{3+}$ defects, respectively[29–31]. Other crystal defects such as $Ce^{3+}$, $Eu^{2+}$, $Cu^{2+}$, $Al-O^--Al$, $Ga^{3+}$, $Sm^{3+}$, $Dy^{3+}$, $Eu^{3+}$, $Tb^{3+}$, $Nd^{3+}$, $Pb^+$, etc. have also been identified using the CL spectra[2,25,32]; however, only a few investigations have been done on the spectral emission region beyond 800 nm in which the principal dosimetric trap emits[31,33,34]. The main source of information on the emission above 800 nm is through radioluminescence obtained from exposure to beta particles. The radioluminescence spectrum of feldspar consists of a broad emission band centered at ~865 nm at room temperature. Major research on this peak was carried out by Trautmann et al.[15], Krbetschek et al.[35], Trautmann et al.[36], Erfurt and Krbetschek[19], and Erfurt[20]. However, this band ~865 nm was initially reported in the CL spectrum by Krbetschek et al.[37]. Based on the comparison of the OSL excitation spectrum (excitation peak ~860 nm) and the IRRL emission spectrum (emission peak ~865 nm) at room temperature, the origin of these two signals was suggested to be from the same trap, i.e. the principal trap[19,35,38,39].

## 3. Cathodoluminescence (CL), x-ray excited optical luminescence (XEOL), and photoluminescence (PL) emission spectra

Figures 1a, b and c show the average CL spectrum, the XEOL and PL emission spectra, respectively, of sample R47, measured at room temperature. The CL spectrum was recorded in the wavelength range of 600-1000 nm and the XEOL spectrum was recorded in the wavelength range of 600-1100 nm. The PL emission spectrum was measured only from 850 to 1100 nm since an 830 nm laser was used to excite the sample, and an 850 nm long pass interference filter was used to detect the NIR emission bands. CL and XEOL emissions were recorded after 48 hours of bleaching in the solar simulator (Hönle SOL 2) in order to empty

the principal trap; this maximizes the radioluminescence signal sensitivity by increasing the concentration of traps available for electron capture.

In each excitation mode, the emission spectrum in the NIR range (850-1100 nm) consists of a broad peak centered at ~900 nm. For the XEOL and PL measurements on the same sample (R47), Kumar et al.[18] showed that this broad peak splits into two peaks centered at ~880 nm (1.41 eV) and ~955 nm (1.30 eV) at a low temperature of 7 K. CL and XEOL mechanisms are similar in terms of ionization and trapping; the only difference is the interaction volume of the crystal. In the case of CL, we are biased by the near-surface traps as the penetration depth of 20 keV electrons is only about 4 µm in feldspar[40]. XEOL reaches a relatively greater penetration depth of about 100 µm using x-rays with a mean energy of ~13 keV[28]. The CL and the XEOL data look very similar in the NIR region indicating that there is no significant modification of the principal trap, closer to the crystal surface. The CL data can be fitted to a linear combination of two Gaussian peaks (inset Figure 1a); the peak maxima are the same as those observed with XEOL and PL at 7 K and 295 K[18]. A peak at ~720 nm (1.72 eV) is also observed in the CL and XEOL emission spectra; this peak has been rigorously studied previously and has been attributed to the Fe-centers[25,30,31,41].

Based on the comparison of CL, XEOL, and PL spectra, we conclude that the CL emission in the NIR region arises from the principal trap as it is similar to the XEOL and PL from the bulk crystal volume. This raises exciting possibilities to map the defects using scanning electron microscope-cathodoluminescence (SEM-CL) imaging. In the remaining paper, we investigate spectrally-resolved CL mapping on 5 different feldspar samples. We henceforth refer to CL in the NIR spectral region as infrared CL (IRCL). The experimental and analytical details are presented in Section 9.

## 4. Mapping of two IRCL emission bands

Figure 2 shows the CL mapping data from sample R47. Figure 2a shows the backscattered electron image (BSE). Figure 2b shows the IRCL intensity map for the feldspar grains; the intensity below a certain value ($I/I_{max} < 0.235$; $I_{max}$=maximum intensity) was masked using a filter (black background). Figure 2c shows the barycenter map of IRCL in the spectral range 880-925 nm, calculated using equation 1. This region was selected to ensure a minimum contamination effect from the adjacent $Fe^{3+}$ peak. We observed certain pixels (very deep red spots), called hot spots, which are likely artifacts from spurious noise during photon detection; therefore the maximum cut-off of 925 nm was imposed on the barycenter value. This resulted in the rejection of only a few pixels per image. A barycenter map reflects a shift in the emission band across pixels. For example, a barycenter is expected to shift to longer wavelengths if peak position or emission band shifts to longer wavelengths. Most of the grains appear to be dominated by the barycenter at longer wavelengths (>915 nm; light and deep reddish regions), and a few grains by slightly shorter wavelengths (<915 nm; yellow-blue-greenish regions). Local CL emission spectra from high (I ≈0.9 $I_{max}$ ; region 1 on Figure 2b) and low intensity (I ≈0.3 $I_{max}$; region 2 on Figure 2b) regions are shown in Figure 2d. Figure 2e shows a map of the relative CL intensity between two bands: band 1 (890-910 nm) and band 2 (960-980nm); the map was created by taking the ratio of areas under the two bands of the local CL emission spectra from each pixel. It can be seen in Figure 2e that the relative intensity between these two bands varies from grain to grain (the blueish region is dominated by band 2 while yellowish and reddish regions are dominated by band 1). It is interesting that the relative intensity between two bands varies even within the grain (Figures 2a, c, and e; circled gains). The ratio map supports the barycenter map: a relative increase in the 955 nm emission band compared to the 880 nm band corresponds to a redshift in the barycenter. Figure 2f shows the elemental map (K/(K+Na)) from EDX. Excluding a couple of grains and grain boundaries with high Na content, the composition generally varies from about 80 to near 100% potassium. It appears that the Na rich regions (bluish regions in Figure 2f) emit a relatively stronger 880 nm IRCL emission (Figures 2c, e), while the K rich regions emit a stronger 955 nm emission.

Figure 3 shows the same data as Figure 2 but for the sample R43. The main differences between Figures 2 and 3 are: 1) the size of each pixel is 12.6 μm for sample R47 and 6.7 μm for sample R43, and 2) I < 0.235 $I_{max}$ and I < 0.135 $I_{max}$ have been treated as background in the case of Figure 2 (sample R47) and Figure 3 (sample R43), respectively.

We extract the following information from the data presented in Figures 2 and 3:

1. The barycenter varies from 880 nm to 925 nm in both samples R47 and 43.

2. From the barycenter map, it is clear that some grains emit at shorter wavelengths (the 880 nm emission band) and some at longer wavelengths (the 955 nm emission band). This suggests that the two IRCL emission bands, and therefore, the relative concentrations of the principal traps responsible for these emissions vary spatially.

3. The barycenter shifts are seen even within an individual K-rich feldspar grain (Figure 2c, circled grain) with a size of ~250 μm. This indicates that the trapping states and thus the luminescence properties can vary within a single-grain at the scale smaller than 250 μm.

4. Na-rich feldspar grains seem to preferentially emit the 880 nm IRCL compared to the K-rich feldspar grains.

## 5. Elemental concentrations and the CL emission bands

Since EDX and CL spectra were collected simultaneously, it is possible to investigate the relationship between chemical composition and luminescence emission for each pixel. We did not observe any significant correlation between the measured elemental concentrations and the CL emission wavelength, except for K, Na, and Fe. Here, we focus on the main cationic constituents K, and Na, as well as Fe. As we are dealing with alkali feldspars only, the Ca concentration in our samples (see Table 1) is not high enough (0.02-0.04%) to attempt any correlations. Elemental ratio K/(K+Na) maps created from the EDX data are shown in Figures 2f and 3f for samples R47 and R43, respectively. It is observed that more than 95% of the elemental map is dominated by the K-content in the case of sample R47 whereas more than 80% of the studied area has K/(K+Na) ratio >0.8 in the case of sample R43. In both samples, most of the grains are highly K-rich (80-100%), however, we occasionally observe significant intra-grain K variation (between 50 to 100%).

The relation between the barycenter in the range 880-925 nm and the K/K+Na of sample R47 ratio is plotted in Figure 4a. Each data point in Figure 4a is an average of two neighboring pixels. Figure 4b shows the same plot for sample R43. We observe a positive correlation between the IRCL emission and the K-content (Pearson's correlation coefficient, r = 0.57 and 0.53 for samples R47 and R43 respectively). The barycenter shifts towards longer wavelengths with increasing K-content. Furthermore, the barycenter is mostly observed at longer wavelengths (>905 nm) for higher K-content (80 to 100% in R47; 75 to 95% in R43). It is also evident that the sample R47 contains relatively less Na-rich pixels compared to the sample R43; the bulk composition of R47 lies between 80-100% K-feldspar.

These data suggest that the IRCL emission at longer wavelengths (for e.g. 955 nm) is more prominent in the K-feldspar end-member regions, whereas lower wavelengths (for e.g. 880 nm) are emitted from the moderately K-rich or Na-rich regions (Figures 4a and b). The interpretation that the longer IRCL band (~955 nm) is emitted preferentially from K feldspar end members is confirmed by our other K-rich samples (see Figures SI 1 and SI 2 for samples R50 and K13, respectively). Elemental maps of these samples are dominated by a K/(K+Na) ratio >0.8 (more than 95% of the data) and the corresponding barycenters are in the range 910-935 nm. Our finds are in the line with those by Thomson et al.[21] and Sellwood et al.[22] who also observed that the IRPL (~955 nm emission) originates preferentially from K-feldspar grains; note that these studies did not detect the 880 nm emission. Thomsen et al.[21] also observed IRSL signals (above room temperature) from both Na- and K-rich regions in spatially resolved luminescence images. Thus, one may conclude that both 880 nm and 955 nm centers contribute to the IRSL (OSL) signal.

The spatial variations in the relative intensities of the two IRCL emissions and their link to K- and Na-content suggest that two different sites are responsible for these emissions. If both the emissions arose from the same site, then one should not see a change in the ratio of the two peaks, irrespective of the spatial location.

In addition to the IRCL emission bands, the CL spectrum shows a broad peak centered at ~720 nm (Figure 1a). This emission band has been suggested to originate from iron ($Fe^{3+}$) substituting $Al^{3+}$ in the tetrahedral sites $T_1$ and/or $T_2$[42–45]. The $Fe^{3+}$ emission peak shifts towards shorter wavelengths (for e.g. 720 nm to 680 nm) when the composition varies from Na-rich to K-rich feldspar[31,46,47]. As reported in the literature, the origin of this shift is suggested to be the change in the Fe-O bond length due to potassium ions in K-rich feldspar and crystal field effect, which modify the energy levels[47]. We created a barycenter map for the $Fe^{3+}$ emission using CL data in the spectral range of 630-800 nm. The shift in $Fe^{3+}$ emission is confirmed in our data, both in the local spectra (Figures 5a and b for samples R47 and R43, respectively) as well as in the correlation plot of barycenter (710 nm to 745 nm) versus the relative K-content (Figures 4c and d for samples R47 and R43, respectively).

## 6. Correlation between $Fe^{3+}$ emission and the IRCL emission bands

We observe that as the $Fe^{3+}$ barycenter shifts towards shorter wavelengths with increasing K-content, the barycenter in the NIR region (IRCL; ~880 and ~955 nm) shifts towards longer wavelengths (Figure 4). These wavelength shifts in the $Fe^{3+}$ and IRCL emissions appear to mirror each other. For sample R47, a barycenter shift for $Fe^{3+}$ emission is observed from 740 nm to 710 nm and a barycenter shift for IRCL emission is observed from 880 nm to 925 nm when the K-content varies from 20 to 100% (Figure 4a and c). For sample R43, the barycenter shift for $Fe^{3+}$ emission is observed from 745 nm to 715 nm and the barycenter shift for IRCL emission is observed from 880 nm to 925 nm when the K-content varies from 20 to 90% (Figure 4b and d).

The relationship between the peak shift of $Fe^{3+}$ and IRCL emissions is confirmed by the local CL spectra. Figures 5a and b show the normalized spectra from regions with different K concentrations for samples R47 and R43, respectively. Figure 5a shows local spectra of the regions with K-contents of about 30%, 90%, 96%, and 98%, for sample R47. Figure 5b shows local spectra of the regions with K-contents of about 30%, 75%, 80%, and 88%, for sample R43. A ~70% change (30% to 98%) in K-content results in about ~45 nm blueshift of the $Fe^{3+}$ peak (~745 nm to ~690 nm) and a ~70 nm redshift of the IRCL peak (~890 nm to ~960 nm) in the case of sample R47. On the other hand, a change of ~60% (30% to 88%) in K-content results in a ~40 nm blueshift of the $Fe^{3+}$ peak (~740 nm to ~700 nm) and a ~70 nm redshift of the IRCL peak (~885 nm to ~955 nm) in the case of sample R43. There is a systematic shift of both the IRCL and the $Fe^{3+}$ peaks with K-content in sample R43 (Figure 5b). There seems to be a tendency that the relative intensity of the IRCL emission decreases (compared to the $Fe^{3+}$ peak) as the $Fe^{3+}$ emission moves to longer wavelengths with a decrease in the K-content. At the same time, there is a peak shift to shorter wavelengths in the IRCL emission. Thus the two peaks tend to come closer, with a simultaneous decrease in the relative IRCL intensity with a decrease in the K-content. A compilation of data from all the local spectra within a sample (Figures 4 and 5) confirms this picture. There exists a negative correlation between barycenters for the IRCL band and for the $Fe^{3+}$ emission; this negative correlation is linked to the K- and Na- content in the measured regions.

To further investigate whether this apparent correlation between $Fe^{3+}$ and IRCL emissions is a coincidence or owing to the common physical mechanism, the same measurements as outlined above were made for a sample (K7) with approximately equal K and Na contents. The data are presented in Figure 6 (note that the size of each pixel in the map is 9.1 µm). Figure 6a shows the backscattered electron image (BSE). Figure 6b shows the barycenter map in the spectral range 755-812 nm; no values were obtained above 810 nm as no emission in the NIR region could be detected. Figure 6c shows the elemental map (K/(K+Na)) from EDX. Local CL emission spectra (800-1000 nm) from low (I ≈0.3 $I_{max}$) and high intensity (I ≈0.9 $I_{max}$) regions are shown in Figure 6d, while spectra in the complete range (600-1000 nm) for regions with different K concentrations are shown in Figure 5c.

The K-content in sample K7 is generally less than 75% (Figure 6c). The $Fe^{3+}$ peak does not shift significantly with a change in K-content; it shifts only by about 735 nm to 730 nm for a change of 20% to 75% K-content in different pixels (Figure 5c). This behavior is reflected in a fairly uniform barycenter across the sample (Figure 6b), which is consistent with earlier CL studies[30]. Interestingly, the IRCL emission bands are generally absent in this sample (Figure 6d). If we compare the K-concentrations, then we should have expected to see the IRCL bands in K7, since medium to low-K regions showed IRCL in R47 and R43. If we compare the absolute concentration of Fe, K7 has about two times the concentration compared to R47 and four times that of R43. It is plausible that a relatively high concentration of $Fe^{3+}$ quenches the IRCL emission when it exists in a moderately K-doped (20-75%) feldspar lattice. In a K-rich lattice such as in sample R50 (K>80%), we observe that IRCL is not quenched (Figure SI 1) even though this sample has a similar absolute concentration of Fe as K7 (Table 1).

We suggest here that the wavelength shifts of IRCL predominantly occurs if the Fe defect is located in the vicinity of the principal trap, forming a possible cluster, in a K-deficient lattice. This leads to competition in electron capture[48] (e.g. $Fe^{4+} + e^- \rightarrow Fe^{3+}$) between the Fe and the principal trap, resulting in an apparent quenching of the IRCL emission. Since the defect responsible for IRCL is unknown, the cause for the shift in the IRCL peak wavelength, and why this shift is in the opposite direction as the $Fe^{3+}$ emission peak, remain to be understood. It is conceivable that the change in the electrostatic field owing to the presence of Fe results in a shift of the IRCL to shorter wavelengths. Further investigations are required to test the proposed cluster or defect-association model.

## 7. Summary and discussion

A combined study of spatially and spectrally resolved cathodoluminescence in the spectral range 800-1000 nm suggests that the relative intensity of the two NIR emission bands reported by Kumar et al.[18] varies spatially within a sample, and even within a grain. We observe a clear correlation between the relative intensities of the IRCL emissions at 880 nm and 955 nm with the K-content in our investigated samples. The IRCL emission at 955 nm appears to be emitted from the samples with high K-concentration. On the other hand, the emission at 880 nm is emitted from the moderately K- or Na-rich samples. This relationship suggests that the principal trap giving rise to the NIR emissions (PL and CL) occurs at two different sites or in two different environments within K- or Na- rich regions. This, in turn, suggests that OSL or IRSL is a mixture of signals originating from principal traps in two different sites; thus, parameters like athermal and thermal stability of OSL or IRSL signals are influenced by the 'mixed' kinetics of these individual sites.

The presence of two defect sites (i.e. two principal traps) has implications for further development of radiation dosimetry using feldspar. Firstly, it has the potential to give information on the defect locations that give rise to OSL/IRSL at the sub-single grain level; this is otherwise not possible by directly measuring the OSL because of its poor efficiency. Secondly, this opens up the potential to develop site-specific dating techniques. Since radioluminescence is a highly sensitive process, we should be able to detect light from very small volumes of the order of $1\text{-}10^3$ µm$^3$; this is less than $10^3\text{-}10^6$ of the total volume of a 100 µm diameter feldspar grain.

We also observe the well-known peak shift of the $Fe^{3+}$ emission with K-content. However, the $Fe^{3+}$ shows a redshift whereas the IRCL shows a blueshift in the barycenter as the K-content decreases. We propose that $Fe^{4+}$ act as a quencher by competing with the principal trap for free electrons. This interpretation implies that the principal trap and $Fe^{3+}$ occur within a defect cluster and the local crystal field altered by the presence of $Fe^{4+}$ results in preferential blue-shifted IRCL emission. This model, however, remains to be tested on structurally and mineralogically well-characterized single crystal samples.

With respect to IR-RF dating[19], our data imply that the bulk signal originates from different sites, with different K and Fe contents. Interference from $Fe^{3+}$ is expected to be higher for the emission band at ~880 nm than that for the emission at ~955 nm. Therefore, both these bands may show different stability and apparent sensitivity change. Our measurements directly demonstrate the potential of spatially resolved IR-RF dating at the sub-grain level, using, for example, a beta source, and a CCD detector with appropriate detection filters for the two IRCL bands. Equally, it may be worthwhile to perform quality control on the single grain measurements by a subsequent compositional analysis such as by µ-XRF; it will be worthwhile to explore how the dose estimates change by selectively including more and more K-rich grains in the analysis (to reduce the signal contamination from $Fe^{3+}$ as the IRCL and the $Fe^{3+}$ peaks are most separated at the highest K concentration.

This study raises the exciting possibility for mapping metastable states in feldspar using optical site-selective techniques such as IRPL and study charge transfer dynamics on spatial scales. The advantage of the PL mapping techniques is that one can avoid contamination from $Fe^{3+}$. On the other hand, one will be restricted by the diffraction limit. Nonetheless, a significant advance may be possible using super-resolution microscopy (SRM) techniques[53].

## 8. Conclusions

For the first time, the spatial distribution of the metastable states (principal trap) in feldspar at a micron-scale (6-22 μm) was studied. We demonstrated that the two NIR cathodoluminescence emission bands (centered at ~880 nm and ~955 nm) vary spatially, even within a single-grain of feldspar. We suggested that these emissions arise from two different defect sites, and hence principal traps, in the feldspar lattice.

We observed a correlation between K-content and the peak positions of the IRCL (principal trap) and $Fe^{3+}$ emissions. We proposed a defect cluster model in which Fe competes with the principal trap to capture free electrons, and hence is responsible for IRCL quenching.

In terms of defect mapping of feldspar, a better spatial resolution in the future may be achieved by reducing the grid size and/or energy of the incident electrons (thereby electron range). Similarly, IRPL confocal microscopy may be explored to yield a spatial resolution of ~0.5 μm; even greater resolution at the scale of tens of nm may be obtainable using the SRM techniques. Measurement of trapping of charge metastable states at such a high-resolution has implications for a) micro/nano dosimetry, b) understanding of luminescence kinetics, and c) developing next-generation luminescence dating techniques based on the sub-single-grain level.

## 9. Experimental and analytical details

We used four $KAlSi_3O_8$ (K-rich) as well as one mixed (K and Na) feldspars. The specifications of the samples (e.g. grain size, provenance, etc.) are summarized in Table 1. Samples are extracted from either sediments or rocks and have been prepared using standard density separation procedure[49,50]. All Samples were bleached for 48 hours in a solar simulator (Hönle SOL 2) before any measurement.

Photoluminescence (PL; excitation using 830 nm laser) and x-ray excited optical luminescence (XEOL; note that XEOL is a form of radioluminescence) were recorded at room temperature (RT; 300 K) using the Risø station for CryOgenic LUminescence Research (COLUR) at DTU Nutech, Denmark. This facility consists of a Horiba spectrofluorometer (Fluorolog-3) modified to include multi-excitation and -detection ports, an x-ray irradiator (filament-based x-ray tube with a tungsten (W) anode, operated at 40 kV anode voltage, 100-µA anode current), and a temperature-controlled closed-loop He cryostat (7-300 K).

All the CL measurements were carried out at the LumiLab, Department of Solid State Sciences, University of Ghent, Belgium, at room temperature (300 K). CL mapping was carried out using a Hitachi S3400-N scanning electron microscope (SEM) operated at 20 kV. The SEM images were obtained using a back-scattered electron (BSE) detector. The secondary electron (SE) detector of this system is not suitable for SEM imaging at the pressure (20 Pa) used here. The cathodoluminescence (CL) spectrum (600-1000 nm) was captured for each pixel in a mapping grid of 128-by-92 pixels using an optical fiber connected to an Acton SP monochromator and ProEM 1600 EMCCD, both from Princeton Instruments. The map's pixel size reflects the spatial resolution that is determined by the chosen grid and the magnification. Magnification was based on the dominant grain size in a sample; it was set to be higher for samples with relatively smaller grains and, therefore, the resolution of the map is different (~6-22 µm) for different samples (see Table 1). The average CL spectrum for the entire sample was calculated using a coarse resolution scanning with the same monochromator and EMCCD.

The electron beam was scanned for a fixed number of times within a pixel, and the data from each pixel is a sum of all these scans. Figure 7 a, b, and c provide an overview of the technique and the CL system. For spectrum analysis, the local spectra (i.e. the CL spectrum measured for each pixel) were corrected for the detector sensitivity, smoothened using Savitzky-Golay filter, and processed in Matlab.

To examine shifts in the emission band or peak position across pixels, we calculated a barycenter of each local emission spectrum in the wavelength range of our interest as follows:

$$\lambda_{bary}\ (nm) = \frac{\int I(\lambda)\cdot \lambda\, d\lambda}{\int I(\lambda)d\lambda}\ , \quad \text{--------------------------------(1)}$$

where $I(\lambda)$ is the CL intensity at a particular wavelength (λ).

The 800-1000 nm emission range was used to evaluate the barycenter of each local spectrum in the NIR spectral region; lower wavelengths were ignored since we were only interested in investigating the principal trap. We did not use absolute CL intensity to derive any conclusions, given the fact that the feldspar grains used in this study do not have a flat, homogeneous surface and absolute intensities thus cannot be compared.

Simultaneous with the CL measurements, major element analysis was carried out using an energy-dispersive x-ray (EDX) spectrometer. A ThermoScientific Noran System 7 enabled elemental mapping and quantification by probing the characteristic x-ray lines of the major elements, Na ($K_\alpha$, 1.0 keV), K ($K_\alpha$, 3.3 keV), Fe ($K_\alpha$, 6.4 keV) Si ($K_\alpha$, 1.7 keV), Al ($K_\alpha$, 1.5 keV), Ca ($K_\alpha$, 3.7 keV), Mg ($K_\alpha$, 1.3 keV), C ($K_\alpha$, 0.3 keV), O ($K_\alpha$, 0.5 keV), and Ba ($L_\alpha$, 4.5 keV) (see Table 1).

## Acknowledgments

We thank Prof. D.J. Huntley, Simon Fraser University, Canada, for samples K7 and K13. We thank David Van Der Heggen, University of Ghent, Belgium, for helping us measure the emission correction factor. We thank Prof. Philippe Smet, University of Ghent, Belgium, for providing us Figure 7 for depicting the CL system.

## Author contributions

R.K. did all the measurements, analyzed the data, and produced the figures with the help of L.M; R.K., M.J., D.P. and L.M. designed the experiments reported here. R.K. and M.J. interpreted the results. R.K. wrote the main draft of the manuscript with support from M.J. D.P. supervised the project and provided lab facilities at the University of Ghent, and M.J. supervised and M.K. co-supervised the project at the Technical University of Denmark. D.V. and J.G. hosted R.K.'s visit at the University of Ghent and provided lab facilities and support for sample preparation. All the authors reviewed and commented on the manuscript. R.K. and M.J. finalized the manuscript with the inputs from the co-authors.

## Competing interests

The authors declare no competing interests.

## Figure captions

Figure 1. (a) Average cathodoluminescence (CL), (b) x-ray excited optical luminescence (XEOL), and (c) photoluminescence (PL) emission spectra of sample R47 at room temperature (~300 K). All the spectra are corrected for their respective system responses; however, no smoothing has been performed. The inset in (a) shows the deconvolution of the CL peak (~900 nm) into two peaks centered at ~880 nm (1.41 eV) and ~955 nm (1.30 eV). CL and XEOL emission spectra were measured after bleaching the sample for 48 hours in a solar simulator. The PL emission spectrum was measured after bleaching followed by 2 hours of x-ray irradiation. These images were generated using Originlab (Version: 8, Url: https://www.originlab.com/).

Figure 2. CL and elemental data (CL-EDX) for sample R47. (a) Backscattered electrons (BSE) image, (b) cathodoluminescence intensity map; the intensity below a normalized value of 0.235 $I_{max}$ has been filtered out using a mask, (c) barycenter map in the range 880-925 nm, (d) local CL spectra of two regions at intensities, I ≈0.9 $I_{max}$ and ≈0.3 $I_{max}$ on the intensity map (square labeled regions 1 and 2 in Figure 2b), (e) relative concentration map of the K-content (K/K+Na), and (f) intensity ratio map between the two NIR bands (890-910 nm and 960-980 nm). Circled grain shows the variation in IRCL emission within the grain. These images were generated using Matlab (Version: R2016, Url: https://www.mathworks.com/products/matlab.html).

Figure 3. CL and elemental data (CL-EDX) for sample R43. (a) BSE image, (b) cathodoluminescence intensity map; intensity below a normalized value of 0.135 $I_{max}$ has been filtered out using a mask, (c) barycenter map in the range 880-925 nm, (d) local CL spectra of two regions at intensities, I ≈0.9 $I_{max}$ and ≈0.3 $I_{max}$ on the intensity map (square labeled regions 1 and 2 in Figure 3b), (e) relative concentration map of the K-content (K/K+Na), and (f) intensity ratio map between the two NIR bands (890-910 nm and 960-980

nm). These images were generated using Matlab (Version: R2016, Url: https://www.mathworks.com/products/matlab.html).

Figure 4. Correlation between the IRCL emission barycenter and relative K concentration in (a) sample R47, (b) sample R43. Correlation between the $Fe^{3+}$ emission barycenter and relative K concentration in (c) sample R47, (d) sample R43. Each data point was obtained by the binning of two nearby pixels. These images were generated using Matlab (Version: R2016, Url: https://www.mathworks.com/products/matlab.html).

Figure 5. Local CL spectra of samples (a) R47, (b) R43, and (c) K7, obtained from low (20-30%), moderate and high K-rich regions (40%-98%). Savitzky-Golay filter was applied to smoothen the spectra. These images were generated using Matlab (Version: R2016, Url: https://www.mathworks.com/products/matlab.html).

Figure 6. CL and elemental data (CL-EDX) for sample K7. (a) BSE image, (b) barycenter map in the range 755-810 nm, (c) relative concentration map of the K-content (K/K+Na), and (d) local CL spectra of two regions at normalized intensities, I ≈0.9 $I_{max}$ and ≈0.3 $I_{max}$ on the intensity map (not shown). These images were generated using Matlab (Version: R2016, Url: https://www.mathworks.com/products/matlab.html).

Figure 7. Schematic representation of the experimental aspects of cathodoluminescence (CL) microscopy. (a) Interaction of electron beam with the sample and the resulting radiations. (b) Sketch of the main components of the scanning electron microscope (SEM) used for the CL mapping. (c) Examples of data generated: the secondary electrons (SE) image, elemental maps, CL maps, and spectra. These images were created using Mircosoft PowerPoint (Version: 2013, Url: https://www.microsoft.com/).

**Table 1:** Compositions for feldspar samples measured using energy-dispersive x-ray spectrometry (EDX); cumulative concentrations (from total mapping area) are given in atom %. Elemental mapping and quantification were done by probing the characteristic x-ray lines of the major elements. Samples R47, R43, and R50 are well characterized and studied previously by Prasad et al.[17]. Samples K7 and K13 are also well characterized and studied previously by Baril[50], Baril and Huntley[51], and Meisl and Huntley[52].

|  | Sample | | | | |
|---|---|---|---|---|---|
|  | R47 | R43 | K7 | R50 | K13 |
| Provenance | Tanzania | China | Colorado, U.S.A | Unkown | Ontario, Canada |

| Type | Sediment | Sediment | Rock | Museum | Rock |
|---|---|---|---|---|---|
| Grain size (µm) | 120-300 | 50-150 | 10-300 | 50-250 | 100-250 |
| CL map resolution (µm) | 12.6 | 6.7 | 9.1 | 11.2 | 22.4 |
| Magnification | 80 | 180 | 110 | 90 | 45 |
| Chemical composition (Atom %) — C | 18.35 ± 0.12 | 26.28 ± 0.13 | 8.86 ± 0.1 | 39.46 ± 0.18 | 39.68 ± 0.18 |
| O | 53.62 ± 0.26 | 50.17 ± 0.24 | 57.98 ± 0.3 | 42.31 ± 0.21 | 41.91 ± 0.2 |
| Na | 0.36± 0.01 | 0.97 ± 0.01 | 3.29 ± 0.03 | 0.71 ± 0.01 | 0.53 ± 0.01 |
| Mg | 0.06 ± 0.01 | 0.07 ± 0.01 | 0.08 ± 0.02 | 0.09 ± 0.01 | 0.07 ± 0.01 |
| Al | 6.1 ± 0.03 | 4.54 ± 0.02 | 7.34 ± 0.04 | 4.45 ± 0.02 | 4.37 ± 0.02 |
| Si | 17 ± 0.06 | 15.25 ± 0.05 | 19.17 ± 0.07 | 10.35 ± 0.03 | 10.61 ± 0.04 |
| K | 4.43± 0.02 | 2.62 ± 0.01 | 3.17 ± 0.03 | 2.53 ± 0.01 | 2.77 ± 0.01 |
| Ca | 0.04 ± 0.01 | 0.04 ± 0.01 | 0.02 ± 0.01 | 0.03 ± 0.01 | 0.02 ±0.01 |
| Fe | 0.02 ± 0.01 | 0.04 ± 0.01 | 0.08 ± 0.01 | 0.09 ± 0.01 | 0.04 ± 0.01 |
| Ba | 0.03 ± 0.01 | 0.02 ± 0.01 | 00 ± 0.01 | 00 ± 0.01 | 0.01 ±0.01 |
| Total | 100 | 100 | 100 | 100 | 100 |

**Figure 1**

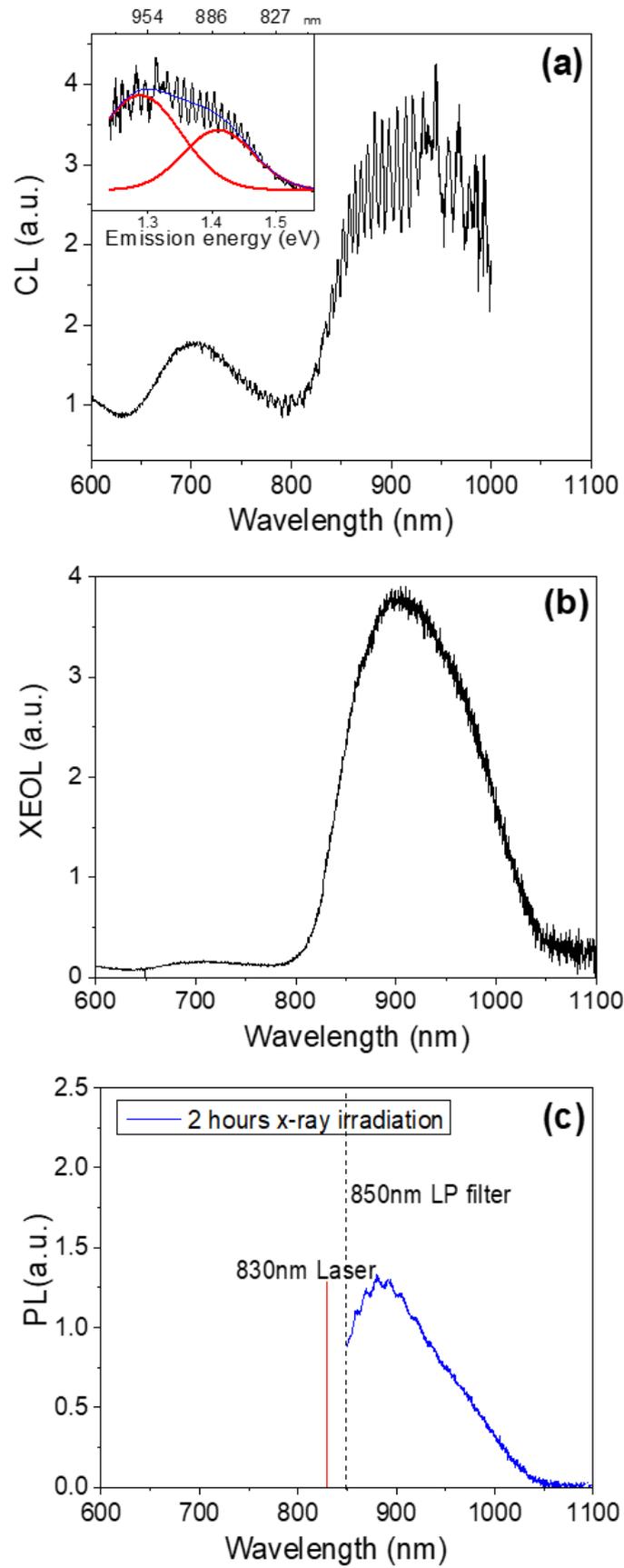

**Figure 2**

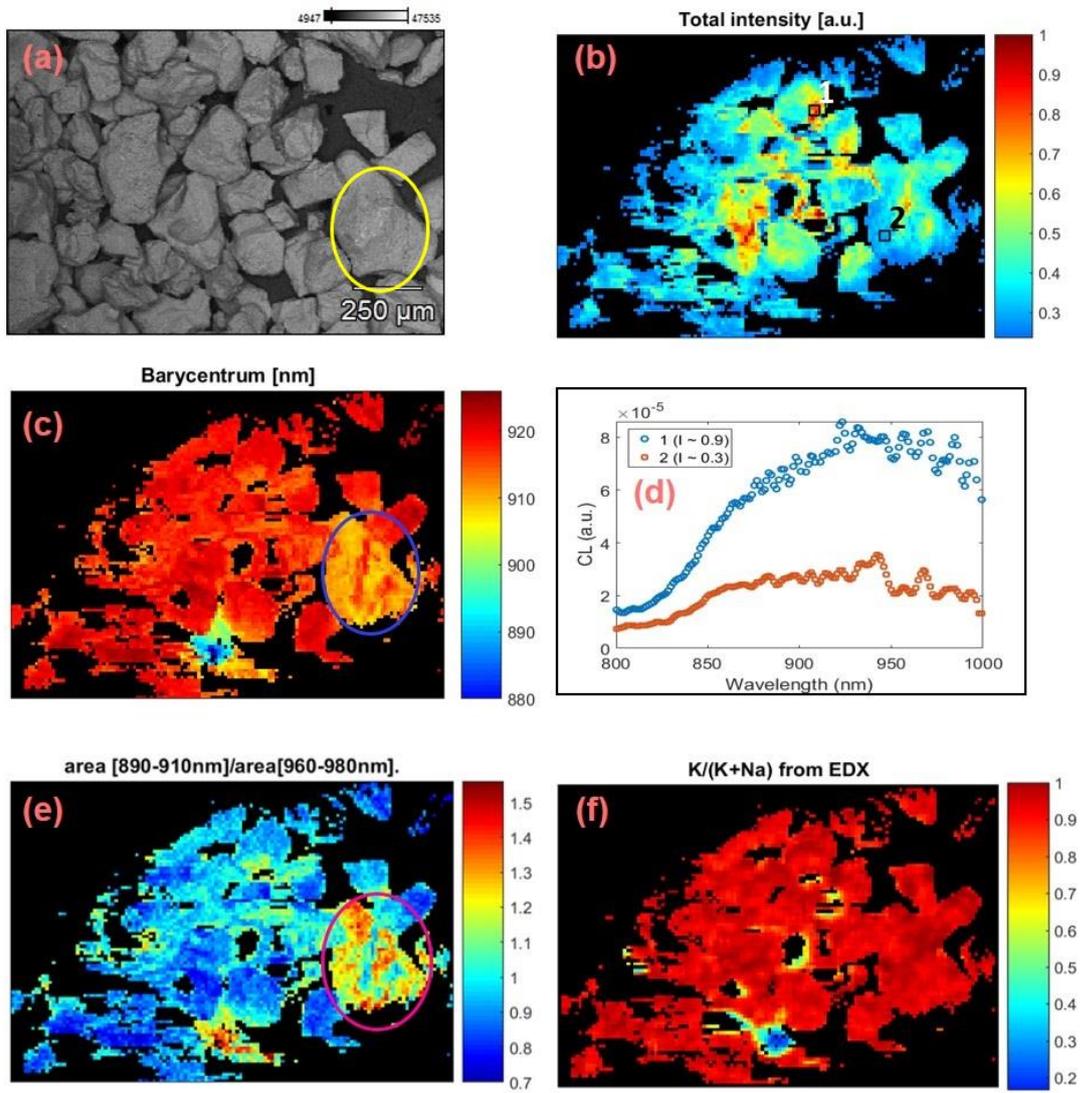

**Figure 3**

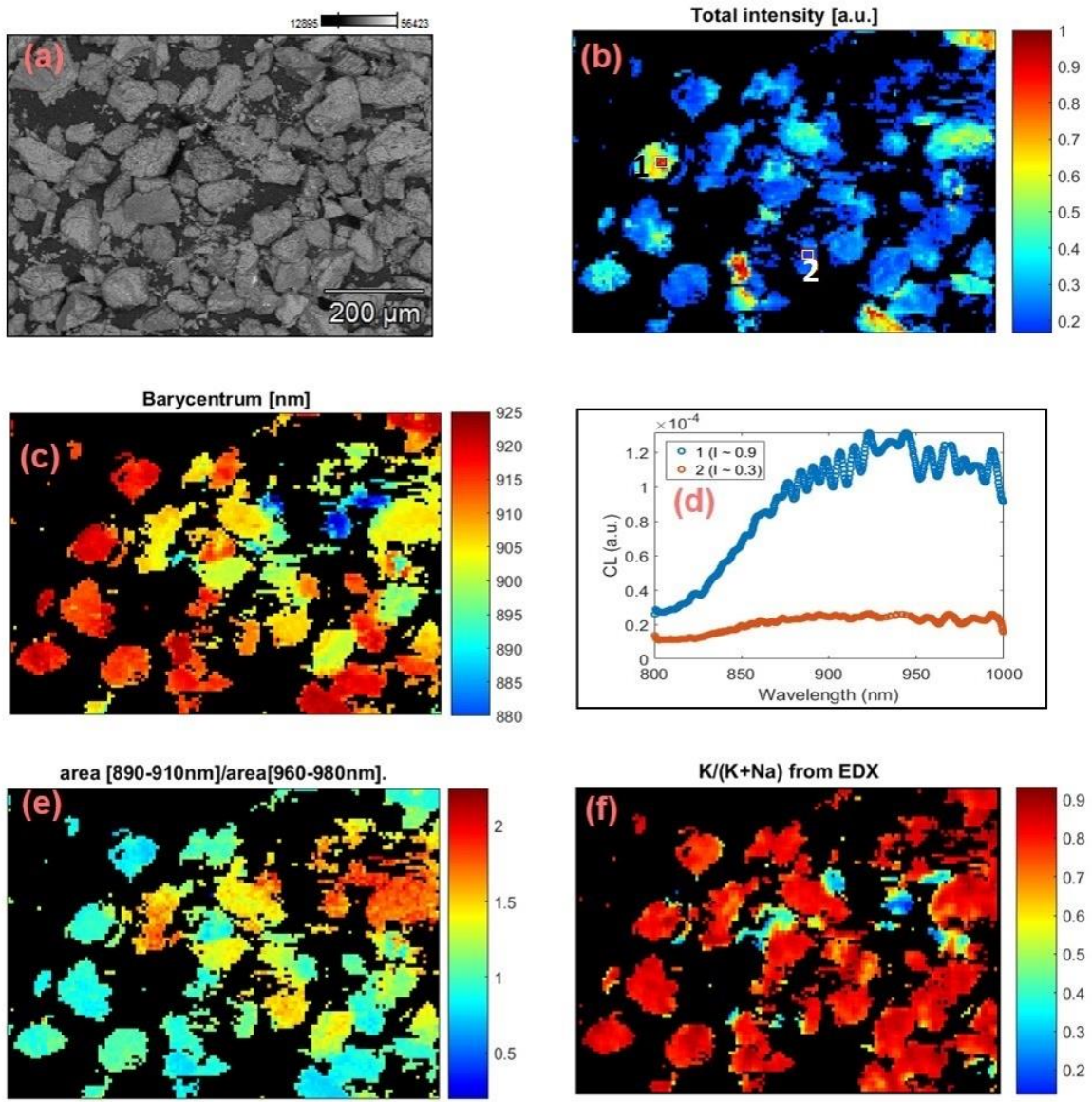

**Figure 4**

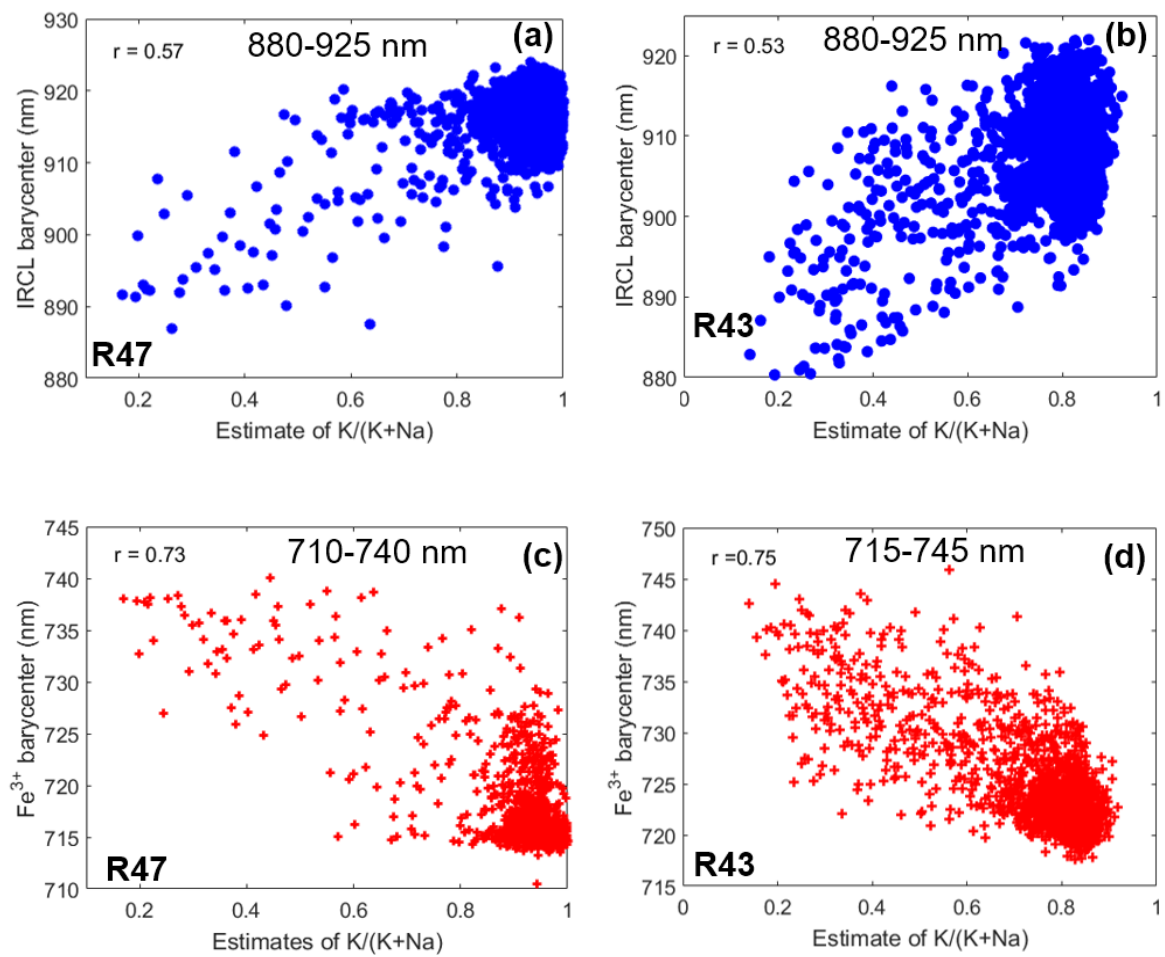

**Figure 5**

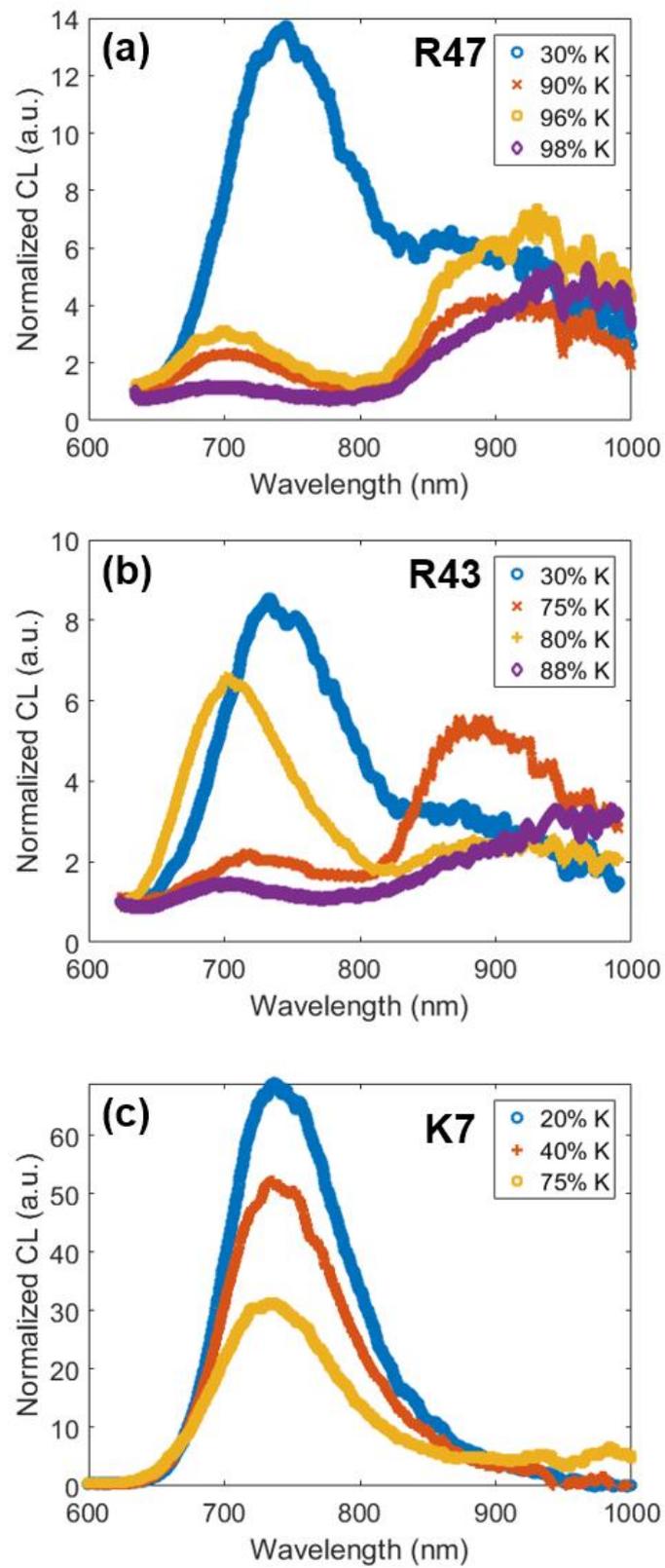

**Figure 6**

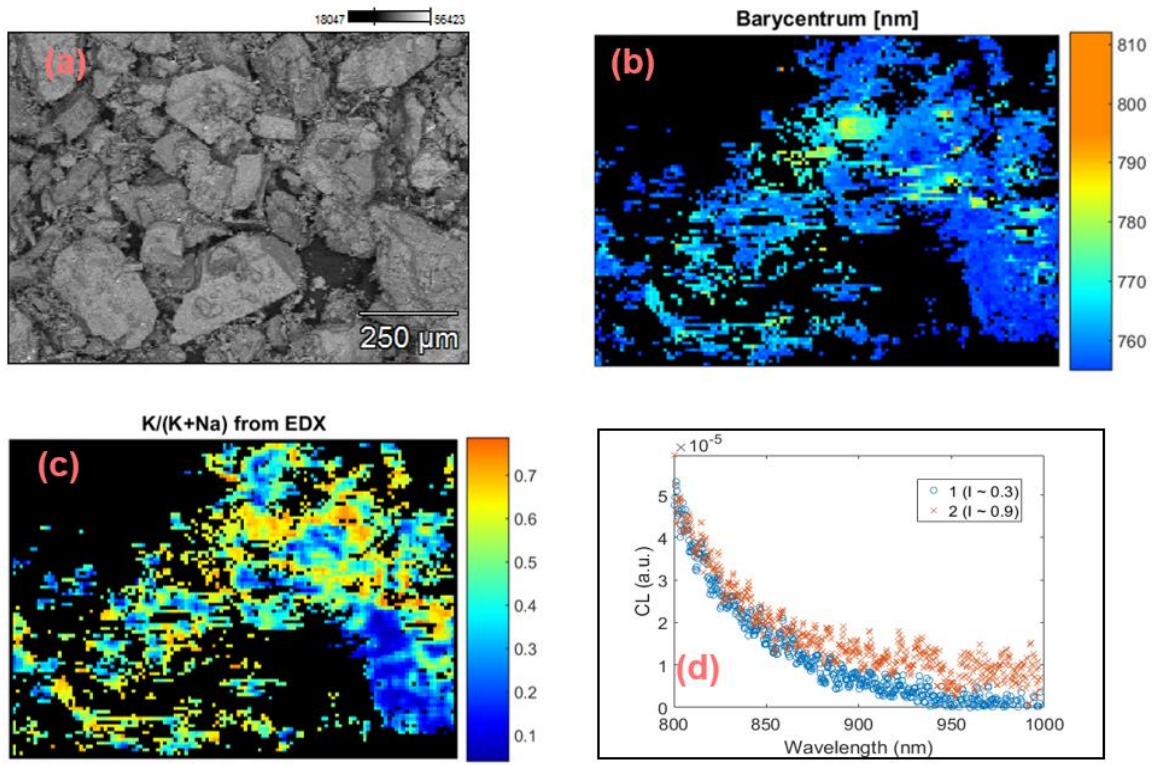

**Figure 7**

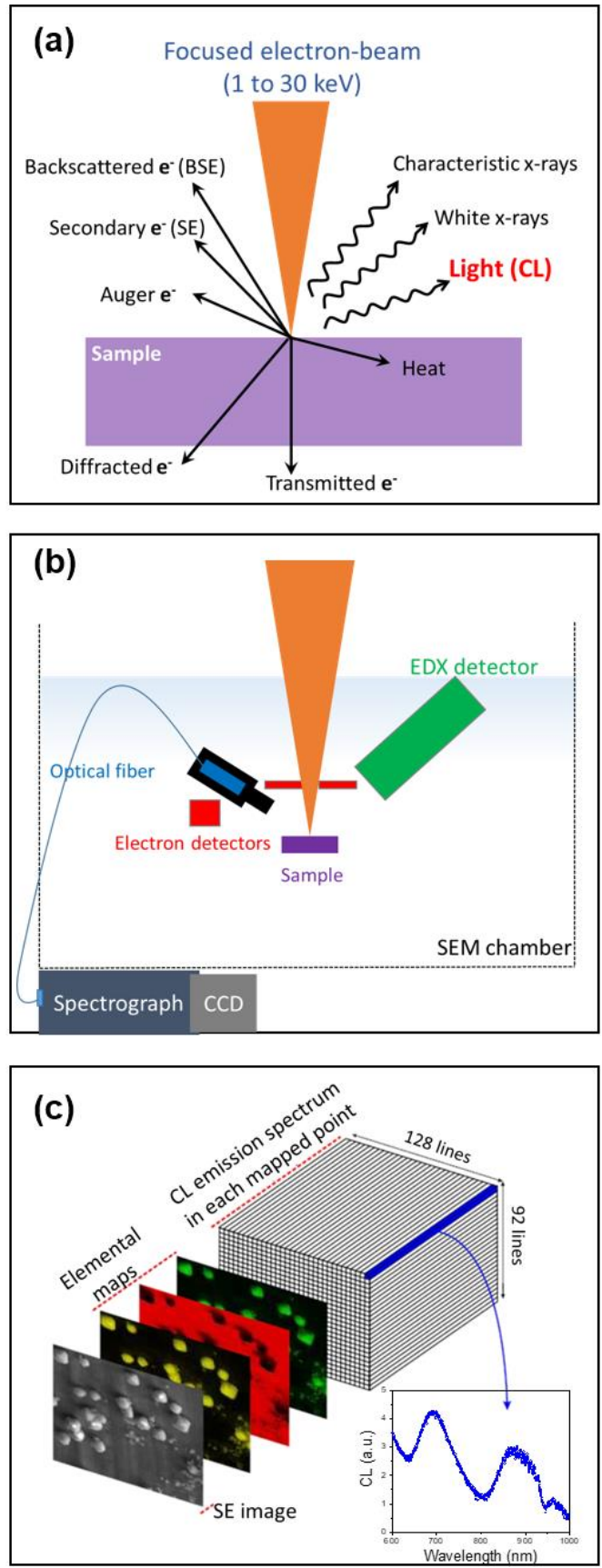